\documentclass[conference]{IEEEtran}
\IEEEoverridecommandlockouts

\usepackage{cite}
\usepackage{amsmath,amssymb,amsfonts}
\usepackage{amsthm}
\usepackage{graphicx}
\usepackage{textcomp}
\usepackage{xcolor}
\usepackage{booktabs}
\usepackage{multirow}
\usepackage{array}
\usepackage{subcaption}
\usepackage[ruled,vlined]{algorithm2e}
\usepackage[margin=0.73in]{geometry}
\providecommand{\nonl}{\relax}

\newtheorem{remark}{Remark}

\begin{document}

\title{Time Window-Based Netload Range Cost Curves for Coordinated Transmission and Distribution Planning Under Uncertainty}

\author{\IEEEauthorblockN{Yujia Li, Alexandre Moreira, Miguel Heleno}
\IEEEauthorblockA{Lawrence Berkeley National Laboratory, Berkeley, CA, USA\\
yujiali@lbl.gov,~AMoreira@lbl.gov,~MiguelHeleno@lbl.gov}}

\maketitle

\begin{abstract}
Mechanisms to coordinate transmission and distribution planning should be regulatory compliant and keep the spheres of DSO and TSO decisions separate, without requiring disclosure of proprietary data or unrealistic computationally expensive T\&D co-simulations. The concept of Netload Range Cost Curves (NRCC) has been recently proposed as simple non-invasive form of coordinating T\&D investments under distribution netload uncertainty. This paper extends the NRCC concept to accommodate the temporal dimension of the T\&D planning process. We propose to compute a hierarchy of certified temporal interface products that represent the different levels of flexibility that distribution networks can provide transmission grids with at the planning stage. The first product (P1) maps distribution investment into scenario-robust, per-window service envelopes within which any TSO service call (to modify load within specified bounds) is guaranteed distribution-network-feasible. The second product (P2) adds lexicographic rebound minimization, preserving P1-optimal service capacity while certifying post-service recovery under three governance variants with qualitatively distinct rebound-budget responses. In our numerical results, based on a real distribution feeder, we compare the performance of our proposed time-window-based flexibility products to an atemporal product (P0) that offers a static bound on the aggregate distribution grid netload across all time periods. Our results demonstrate the superiority of our proposed products in properly valuing the benefits of incremental investments in storage to allow for temporal flexibility. 
\end{abstract}


\section*{Nomenclature}
\begin{IEEEdescription}[\IEEEsetlabelwidth{$\boldsymbol{u}\in\{0,1\}^{n_u}$}]
\item[\textit{Indices and Sets}]
\item[$k\in\mathcal{K}$] Budget tier index and set.
\item[$t\in\mathcal{T}$] Time-step index; $\mathcal{T}$ is the full planning horizon.
\item[$s\in\mathcal{S}$] distributed generation-growth scenario index and scenario set.
\item[$\nu\in\mathcal{W}$] Service-window index and window set.
\item[$\mathcal{T}(\nu)$] Time steps belonging to service window $\nu$.
\item[$\mathcal{R}(\nu)$] Rebound window for P2-b.
\item[$\mathcal{P}(\nu)$] Protected window for P2-a.
\item[$\mathcal{N}^{\rm sub}$] T\&D boundary substation nodes.
\item[$\mathcal{Q}$] Product-variant index set; $\mathcal{Q}=\{{\rm a},{\rm b},{\rm c}\}$.
\item[$\Omega_\nu$] Service-call set for window $\nu$; see~\eqref{eq:callset}.
\item[$\widetilde{\Omega}_\nu$] Stress-screening subset $\widetilde{\Omega}_\nu\subset\Omega_\nu$ used in MILP approximation.
\item[$\mathcal{X}^{\rm pl}$] Planning feasibility set.
\item[$\mathcal{X}^{\rm op}_s(\cdot)$] Operational feasibility set under scenario $s$.
\item[$\mathcal{X}^{\rm op}_{s,\nu}(\cdot)$] Service-augmented operational set; see~\eqref{eq:M3}.
\item[\textit{Parameters}]
\item[$\Gamma^0$] Least-cost feasible investment budget.
\item[$\Gamma_k$] Total investment budget at tier $k$; $\Gamma_k=\Gamma^0+\Delta\Gamma_k$.
\item[$\Delta\Gamma_k$] NRCC incentive increment offered at tier $k$.
\item[$W$] Weighting factor between direct and reverse peak objectives.
\item[$\Lambda^D,\Lambda^R$] Expected peak direct / reverse netloads.
\item[$\lambda^{D}_k$] P0-optimal direct peak cap at tier $k$.
\item[$\bar{p}^0_{t,s}$] Baseline substation netload under scenario $s$ at time $t$.
\item[$\rho_\nu$] Importance weight assigned to service window $\nu$.
\item[$\beta_\nu^{\downarrow},\beta_\nu^{\uparrow}$] Direction scoring factors for downward / upward service.
\item[$\theta_\nu^{\downarrow},\theta_\nu^{\uparrow}$] Sustained duration multipliers [h]; $E_\nu^{\downarrow}=\theta_\nu^{\downarrow}R_\nu^{\downarrow}$, $E_\nu^{\uparrow}=\theta_\nu^{\uparrow}R_\nu^{\uparrow}$.
\item[$R^{\star\downarrow}_{\nu,k},R^{\star\uparrow}_{\nu,k}$] P1-optimal power ratings for window $\nu$ at tier $k$.
\item[$E^{\star\downarrow}_{\nu,k},E^{\star\uparrow}_{\nu,k}$] P1-optimal energy budgets; $E^{\star\downarrow}_{\nu,k}=\theta_\nu^{\downarrow}R^{\star\downarrow}_{\nu,k}$.
\item[$\Delta t$] Time step duration.
\item[\textit{Decision Variables}]
\item[$\boldsymbol{u}\in\{0,1\}^{n_u}$] Binary investment decision vector.
\item[$\boldsymbol{v}\in\mathbb{R}_+^{n_v}$] Continuous sizing decision vector.
\item[$\boldsymbol{w}_s$] Operational variable vector under scenario $s$.
\item[$p^{\rm sub}_{t,s}$] Substation netload at time $t$ under scenario $s$.
\item[$F^{\rm DN}(\cdot)$] Annualized total DN cost function. 
\item[$\lambda^D,\lambda^R$] Peak direct / reverse substation netload bounds.
\item[$R_\nu^{\downarrow},R_\nu^{\uparrow}$] Downward / upward power rating for window $\nu$.
\item[$E_\nu^{\downarrow},E_\nu^{\uparrow}$] Downward / upward energy budget for window $\nu$.
\item[$\xi^{\downarrow}_t,\xi^{\uparrow}_t$] Downward / upward service-call trajectory at time $t$.
\item[$\eta$] Worst-case peak rebound metric. 
\end{IEEEdescription}


\section{Introduction}
Modern power systems planning is evolving from the classic approach of treating the transmission network in isolation. Behind-the-meter distributed generation are rapidly reshaping the netload profiles seen at transmission and distribution (T\&D) boundary substations, altering both their magnitude and temporal shape \cite{9395702}. Since this boundary netload is governed by distribution-side demand, local constraints, and investment decisions of distribution generations, transmission upgrades designed under fixed boundary assumptions will likely lead to either overly conservative or insufficient across plausible distributed generation and load growth pathways \cite{10982438}.

This coupling creates a practical need for coordinated planning across T\&D. In principle, distribution planners can invest in local upgrades and controllable resources to reshape boundary netload trajectories, potentially deferring transmission reinforcements \cite{10982438}. In practice, coordination is difficult because TSOs and DSOs operate under different objectives, investment cycles, model granularities, and data-access constraints \cite{11271528}. As a result, from a DSO perspective, it remains challenging to provide the transmission planner with netload interface that is simultaneously decision-relevant for transmission investment and credible with respect to distribution feasibility. In addition, to make T\&D planning coordination realistic and regulatory compliant, it is important to keep the spheres of DSO and TSO decisions separate, without requiring disclosure of proprietary data or computationally expensive T\&D co-simulations.

To address this challenge, our prior work introduced the concept of netload range cost curves (NRCCs) as a compact, non-intrusive planning interface  for coordinated T\&D planning under distributed generation growth uncertainty \cite{11271528}. The NRCC framework can be seen as a product offered by the DSO to the TSO, in which the DSO provides a menu that pairs peak substation net-load guarantees with the corresponding distribution upgrade costs. This enables the transmission planner to trade off transmission reinforcements against distribution-side investments while keeping distribution network models and internal constraints local to the DSO.


However, the current NRCC framework is solely focused on providing peak load guarantees at the substation, which presents three main limitations for T\&D planning: first, a bound on the distribution absolute peak cannot capture the temporal structure of transmission stress (i.e., whether that peak is reduced at a time that is critical for the transmission infrastructure); second, when flexibility is provided by energy-limited resources, such as battery storage, there is an energy payback period that is not explicitly allocated to periods that are more convenient for transmission planning; third, the existing framework offers no mechanism for the DSO to translate distribution investment into time-specific flexibility commitments that are simultaneously certifiable by the TSO and consistent with internal network constraints.



This paper extends the NRCC framework from a peak-only menu to a hierarchy of certified temporal interface products that explicitly characterize both service delivery and outside-window rebound behaviour. The contributions are threefold:
(i) We formulate a service-window product (P1) to map a distribution investment budget into per-window power and energy ratings within which any transmission service call is guaranteed distribution-network-feasible under all distributed generation-growth scenarios, moving the T\&D interface from a single peak guarantee to a scenario-robust flexibility envelope that the TSO can query window by window; 
(ii) We introduce a lexicographic rebound-minimization stage (P2) that preserves P1-optimal service capacity exactly while minimizing worst-case outside-window peak rebound under alternative rules.

The remainder of this paper is organized as follows. 
Section~II defines the temporal interface product hierarchy and the associated optimization models. Section~III describes the solution workflow. Section~IV presents numerical results and discusses implications for flexibility procurement across T\&D at the planning stage. 
Section~V concludes the paper and outlines future directions.

\section{Temporal Interface Products}

The NRCC framework introduced in \cite{11271528} involves an atemporal product P0 that offers (from the DSO to the TSO) different pairs of distribution investment budgets and corresponding peak net load at the point of connection between distribution and transmission grids. This paper extends this framework to a hierarchy of temporal interface products, summarized in Table~\ref{tab:products}. P1 augments P0 with a per-window service envelope defined by a maximum deliverable power adjustment and an energy budget, within which any TSO service call is guaranteed distribution-network-feasible under all distributed generation-growth scenarios. P2 locks P1-optimal service level and further manages post-service rebound.

\begin{table}[t]
\centering
\caption{Interface product hierarchy.}
\label{tab:products}
\small
\begin{tabular}{@{}p{0.7cm}p{4.6cm}p{1.4cm}p{0.6cm}@{}}
\toprule
\textbf{Prod.} & \textbf{Interface guarantee} 
& \textbf{Scope} & \textbf{Model} \\
\midrule
P0 & Peak/valley bounds per budget tier 
   & $\mathcal{T}$ & 2 \\
P1 & Per-window power/energy adjustments; 
     any TSO call is DN-feasible under all scenarios
   & $\mathcal{T}(\nu)$ & 3 \\
P2 & P1 ratings locked; rebound minimized;
     variants a/b/c differ in governance scope
   & variant-dependent & 4 \\
\bottomrule
\end{tabular}
\end{table}


\subsection{Baseline Models in the Original NRCC Framework}

\subsubsection{Model 1: Scenario-Based Least-Cost Distribution Network Planning (LCDNP)}
Model~1 selects the minimum-cost distribution network (DN) investment plan that remains operationally
feasible under all distributed generation growth scenarios. 
\begin{subequations}\label{eq:M1}
\begin{align}
\Gamma^0 = \min_{\boldsymbol{u},\boldsymbol{v},\boldsymbol{w}_s}\quad
& F^{\rm DN}(\boldsymbol{u},\boldsymbol{v},\{\boldsymbol{w}_s\}_{s\in\mathcal{S}}) \label{eq:M1_obj}\\
\text{s.t.}\quad
& (\boldsymbol{u},\boldsymbol{v}) \in \mathcal{X}^{\rm pl}, \label{eq:M1_pl}\\
& \boldsymbol{w}_s \in \mathcal{X}^{\rm op}_s(\boldsymbol{u},\boldsymbol{v}), \quad \forall s\in\mathcal{S}, \label{eq:M1_op}\\
& \boldsymbol{u}\in\{0,1\}^{n_u},\quad \boldsymbol{v}\in\mathbb{R}_+^{n_v}. \label{eq:M1_var}
\end{align}
\end{subequations}
Here $\boldsymbol{u}$ contains binary investment decisions, e.g., line reconductoring, voltage regulator and transformer placement,  and $\boldsymbol{v}$ contains continuous sizing variables, e.g., storage capacity. Vector $\boldsymbol{w}_s$ collects all operational variables under scenario $s\in\mathcal{S}$. The objective $F^{\rm DN}$ comprises annualized capital investment costs plus a penalty for load shedding, and its optimal value, $\Gamma^0$, is referred to as the \emph{least-cost feasible budget}. Let $\bar{p}^0_{t,s}$ denote the substation netload at time $t$ under scenario $s$ in the Model~1 optimal operating schedule $\{\boldsymbol{w}^0_s\}$; this scenario-specific baseline profile is used as the reference for service-window adjustments in Models~3 and~4.

\subsubsection{Model 2: Transmission-Aware Distribution Network Planning (TADNP) (P0)}
Given an NRCC incentive $\Delta\Gamma_k$ from the TSO, Model~2 minimizes the worst-case peak netload range at boundary substations across all distributed generation growth scenarios, subject to a total investment budget of
$\Gamma^0+\Delta\Gamma_k$:
\begin{subequations}\label{eq:M2}
\begin{align}
\min_{\boldsymbol{u},\boldsymbol{v},\boldsymbol{w}_s,\lambda^D,\lambda^R}\quad
& W[\lambda^D\!-\!\Lambda^D]^+\!+\!(1\!-\!W)[\lambda^R\!-\!\Lambda^R]^+ \label{eq:M2_obj}\\
\text{s.t.}\quad
& \eqref{eq:M1_pl},\,\eqref{eq:M1_op},\,\eqref{eq:M1_var}, \label{eq:M2_inherit}\\
& {-\lambda^R \le p^{\rm sub}_{n,t,s} \le \lambda^D},\notag\\
&\hspace{20pt}\forall n\in\mathcal{N}^{\rm sub},\,t\in\mathcal{T},\,s\in\mathcal{S}, \label{eq:M2_bound}\\
& \lambda^D,\lambda^R \ge 0, \label{eq:M2_pos}\\
& F^{\rm DN}(\boldsymbol{u},\boldsymbol{v},\{\boldsymbol{w}_s\}) \le \Gamma^0+\Delta\Gamma_k. \label{eq:M2_budget}
\end{align}
\end{subequations}
Here $[\cdot]^+=\max\{\cdot,0\}$, $W\in[0,1]$ is a weighting factor, and $\Lambda^D,\Lambda^R$ are the expected peak direct/reverse netloads from the deterministic LCDNP under the expected scenario. Constraint~\eqref{eq:M2_bound} enforces that the worst-case peak direct and reverse substation netloads across all scenarios do not exceed $\lambda^D$ and $\lambda^R$, which drive transmission-side reinforcement needs. Solving~\eqref{eq:M2} for each $k\in\mathcal{K}$ yields one P0 NRCC menu entry: the pair $(\Delta\Gamma_k,\,[\lambda^R_k,\lambda^D_k])$.

\begin{remark}
The P0 menu should be considered by the transmission planner through an NRCC-informed transmission expansion model, one of $|\mathcal{K}|$ DSO service options can be selected at each boundary substation. That model will optimize the least-cost combination of transmission investments and NRCC purchases across all boundary nodes. The optimization of the transmission planner decision making is out of the scope of this paper and we focus exclusively on the DSO-side product construction.
\end{remark}

\subsection{Temporal NRCC Extension: Service-Window Products and Rebound Management}

\subsubsection{Service-Window Call Set}
A \emph{service call} represents the boundary netload adjustment trajectory that the transmission planner selects within the published service envelope during window $\nu\in\mathcal{W}$. Let $\mathcal{T}(\nu)$ denote the set of time periods in window $\nu$. The admissible service call set is:
\begin{align}
\Omega_\nu(\boldsymbol{R},\,\boldsymbol{E}) := \Big\{&(\xi^{\downarrow}_t,\,\xi^{\uparrow}_t)_{t\in\mathcal{T}(\nu)} \;\ \Big|
  \quad 0\le \xi^{\downarrow}_t \le R_\nu^{\downarrow},\notag\\
  &\quad {\textstyle\sum_{t}} \xi^{\downarrow}_t\Delta t \le E_\nu^{\downarrow}, \quad 0\le \xi^{\uparrow}_t \le R_\nu^{\uparrow},\notag\\
  &\quad {\textstyle\sum_{t}} \xi^{\uparrow}_t\Delta t \le E_\nu^{\uparrow}\Big\}. \label{eq:callset}
\end{align}
Here $\xi^\downarrow_t$ and $\xi^\uparrow_t$ denote the downward and upward boundary power adjustments at time $t$, corresponding to load reduction or local resources' extra injection/consumption during the service window. The energy budget is parameterized by a sustained duration multiplier: 
$E^\downarrow_\nu = \theta^\downarrow_\nu R^\downarrow_\nu$ and $E^\uparrow_\nu = \theta^\uparrow_\nu R^\uparrow_\nu$, where $\theta^\downarrow_\nu, \theta^\uparrow_\nu \geq 0$ represent the maximum number of hours the resource can sustain full-power delivery, reducing $\Omega_\nu$ to a family indexed by the maximum deliverable adjustments $(R^\downarrow_\nu, R^\uparrow_\nu)$ alone.

\subsubsection{Model 3: Temporal NRCC Product Design (P1)}
Model~3 co-designs distribution investments and service envelope parameters at budget tier $k$:
\begin{subequations}\label{eq:M3}
\begin{align}
&\max_{\boldsymbol{u},\boldsymbol{v},\boldsymbol{w}_s,\,\boldsymbol{R},\,\boldsymbol{E}} \quad
\sum_{\nu\in\mathcal{W}} \rho_\nu\Big(\beta_\nu^{\downarrow} R_\nu^{\downarrow} + \beta_\nu^{\uparrow} R_\nu^{\uparrow}\Big) \label{eq:M3_obj}
\\
\text{s.t.}\quad
& (\boldsymbol{u},\boldsymbol{v}) \in \mathcal{X}^{\rm pl},\quad \label{eq:M3_planning}\\ 
&F^{\rm DN}(\boldsymbol{u},\boldsymbol{v},\{\boldsymbol{w}_s\}) \le \Gamma^0+\Delta\Gamma_k, \label{eq:M3_budget}\\
& \boldsymbol{u}\in\{0,1\}^{n_u},\quad \boldsymbol{v}\in\mathbb{R}_+^{n_v}. \label{eq:M3_var}\\
& \forall \nu\in\mathcal{W},\ s\in\mathcal{S},\ (\boldsymbol{\xi}^{\downarrow},\boldsymbol{\xi}^{\uparrow}) \in\Omega_\nu(\boldsymbol{R},\,\boldsymbol{E}): \nonumber\\
&\quad \exists\; \boldsymbol{w}_s\in
  \mathcal{X}^{\rm op}_{s,\nu}\!\big(\boldsymbol{u},\boldsymbol{v};\boldsymbol{\xi}^{\downarrow},\boldsymbol{\xi}^{\uparrow}\big), \label{eq:M3_deliver}
\end{align}
\end{subequations}
where:
\begin{align}
    &\mathcal{X}^{\rm op}_{s,\nu}(\cdot):=\left\{\boldsymbol{w}_s:\;
\begin{array}{@{}l@{}}
  \boldsymbol{w}_s\in\mathcal{X}^{\rm op}_s(\boldsymbol{u},\boldsymbol{v}),\\[1pt]
  p^{\rm sub}_{t,s}=\bar{p}^0_{t,s}-\xi^{\downarrow}_t+\xi^{\uparrow}_t,\;
  \forall t\in\mathcal{T}(\nu),\\[1pt]
  -\lambda^{R}_k\le p^{\rm sub}_{t,s}\le\lambda^{D}_k,\;
  \forall t\in\mathcal{T}\setminus\mathcal{T}(\nu)
\end{array}\right\}.\nonumber
\end{align}

The objective~\eqref{eq:M3_obj} maximizes the weighted sum of deliverable adjustments across all windows. 
Constraints~\eqref{eq:M3_planning}--\eqref{eq:M3_budget} enforce planning feasibility and the investment budget. 
Constraint~\eqref{eq:M3_deliver} is the \emph{deliverability} requirement: for every admissible service call and every scenario, there must exist a DN-feasible operating schedule implementing the requested boundary adjustment. 
Outside the service window, $\mathcal{X}^{\rm op}_{s,\nu}$  enforces $-\lambda^{R}_k \le p^{\rm sub}_{t,s} \le \lambda^{D}_k$, where $\lambda^{D}_k$ and $\lambda^{R}_k$ are the P0-optimal bounds at tier $k$, ensuring that the P1 product does not worsen outside-window transmission conditions relative to the P0 baseline. 
In its exact form, constraint~\eqref{eq:M3_deliver} requires verification over all calls in $\Omega_\nu$. 

\subsubsection{Model 4: Rebound-Aware Refinement (P2)}
For energy-limited flexibility resources, service within $\mathcal{T}(\nu)$ induces energy recovery outside the window that can create uncontrolled netload excursions if left ungoverned. Model~4 performs a lexicographic second stage given the P1-optimal deliverable adjustments and energy budgets $(R^{\star\downarrow}_{\nu,k}, R^{\star\uparrow}_{\nu,k}, E^{\star\downarrow}_{\nu,k}, E^{\star\uparrow}_{\nu,k})$ for investment budget tier $k$ from Model~3: it preserves each per-window rating at its P1-optimal level and minimizes the worst-case peak rebound $\eta$:

\begin{subequations}\label{eq:M4}
\begin{align}
\min_{\boldsymbol{u},\boldsymbol{v},\boldsymbol{w}_s,\,\boldsymbol{R},\,\boldsymbol{E},\,\eta}\quad
& \eta \label{eq:M4_obj}\\
\text{s.t.}\quad
& F^{\rm DN}(\boldsymbol{u},\boldsymbol{v},\{\boldsymbol{w}_s\}) \le \Gamma^0+\Delta\Gamma_k, \label{eq:M4_budget}\\
& \boldsymbol{u}\in\{0,1\}^{n_u},\quad \boldsymbol{v}\in\mathbb{R}_+^{n_v}. \label{eq:M4_var}\\
& R_\nu^\downarrow \ge R^{\star\downarrow}_{\nu,k},\quad R_\nu^\uparrow \ge R^{\star\uparrow}_{\nu,k},\nonumber\\
& E_\nu^\downarrow \ge E^{\star\downarrow}_{\nu,k},\quad E_\nu^\uparrow \ge E^{\star\uparrow}_{\nu,k},\quad \forall \nu\in\mathcal{W}, \label{eq:M4_lock}\\
& \forall \nu\in\mathcal{W},\; s\in\mathcal{S},\; (\boldsymbol{\xi}^{\downarrow},\boldsymbol{\xi}^{\uparrow})\in \Omega_\nu(\boldsymbol{R},\,\boldsymbol{E}): \nonumber\\
&\quad \exists\; \boldsymbol{w}_s\in \mathcal{Y}^q_{s,\nu}\!\big(\boldsymbol{u},\boldsymbol{v};\boldsymbol{\xi}^{\downarrow},\boldsymbol{\xi}^{\uparrow};\eta\big). \label{eq:M4_deliver}
\end{align}
\end{subequations}
Here $q\in\{{\rm a},{\rm b},{\rm c}\}$ indexes the product variant ($q={\rm c}$ is a special case of ${\rm a}$; see below) and $\mathcal{Y}^q_{s,\nu}$ is the rebound-augmented recourse set. Constraint~\eqref{eq:M4_lock} is the \emph{P1-capacity floor}: it pins the minimum deliverable adjustments and energy budgets at their P1-optimal values so that Model~4 cannot sacrifice service capacity in pursuit of lower rebound. Critically, Model~4 re-optimizes all investments from scratch under the same budget $\Gamma_k$, inheriting only the service level from Model~3; this allows a different resource portfolio to be selected if it achieves lower rebound at the same service level. Constraint~\eqref{eq:M4_deliver} requires, for every admissible service call and every scenario, the existence of a DN-feasible schedule that implements the boundary adjustment and obeys the outside-window governance rule of variant $q$. 
\textbf{P2-a} (protected window $\mathcal{P}(\nu)$, $q={\rm a}$) defines:
\begin{align}
\mathcal{Y}^{\rm a}_{s,\nu}&(\cdot) := \Bigl\{ \boldsymbol{w}_s\in\mathcal{X}^{\rm op}_{s,\nu}(\cdot) : \bigl|p^{\rm sub}_{t,s}-\bar{p}^0_{t,s}\bigr|\le\eta,\;\forall t\in\mathcal{P}(\nu),\notag\\
&-\lambda^{R}_k \le p^{\rm sub}_{t,s}\le\lambda^{D}_k,\;\forall t\in\mathcal{T}\setminus(\mathcal{T}(\nu)\cup\mathcal{P}(\nu)) \Bigr\}. \tag{5e} \label{eq:M4_outa}
\end{align}
\textbf{P2-b} (rebound window $\mathcal{R}(\nu)$, $q={\rm b}$) defines:
\begin{align}
\mathcal{Y}^{\rm b}_{s,\nu}&(\cdot) := \Bigl\{ \boldsymbol{w}_s\in\mathcal{X}^{\rm op}_{s,\nu}(\cdot) : \bigl|p^{\rm sub}_{t,s}-\bar{p}^0_{t,s}\bigr|\le\eta,\;\forall t\in\mathcal{R}(\nu),\notag\\
&p^{\rm sub}_{t,s}=\bar{p}^0_{t,s},\;\forall t\in\mathcal{T}\setminus(\mathcal{T}(\nu)\cup\mathcal{R}(\nu)) \Bigr\}. \tag{5f} \label{eq:M4_outb}
\end{align}
Constraints~\eqref{eq:M4_obj}--\eqref{eq:M4_deliver} 
using $\mathcal{Y}^{\rm a}_{s,\nu}$ and 
$\mathcal{Y}^{\rm b}_{s,\nu}$ constitute 
\textbf{Model~4-a} and \textbf{Model~4-b}, respectively.

The two variants reflect complementary positions along a 
flexibility-predictability trade-off. \textbf{P2-a} 
designates $\mathcal{P}(\nu)$ as a \emph{protected window} 
where post-service energy recovery is bounded by $\eta$, while hours outside $\mathcal{T}(\nu)\cup\mathcal{P}(\nu)$ retain the P1 peak-cap guarantee. This suits applications 
where the TSO must avoid new peak violations following service but can tolerate P1-level behavior elsewhere. 
On the other hand, \textbf{P2-b} designates $\mathcal{R}(\nu)$ as the \emph{sole} window where energy recovery may deviate from baseline, locking all remaining hours to the baseline netload exactly and providing the TSO with full boundary predictability outside $\mathcal{T}(\nu)\cup\mathcal{R}(\nu)$, at the cost of a potentially larger $\eta$ within that slot. 
Note that P2-a subsumes a global rebound envelope as a special case when $\mathcal{P}(\nu)=\mathcal{T}
\setminus\mathcal{T}(\nu)$; this variant, denoted \textbf{P2-c}, is evaluated in the case study (Section~\ref{sec:results}).

\begin{remark}
Unlike P0, which publishes a single peak bound per 
substation, P1 and P2 certify a window-specific 
flexibility envelope within which the TSO can freely 
shape its service requests according to transmission-level 
needs. This supports more targeted procurement, e.g., 
calling service only during coincident stress hours, and 
can be integrated into TSO-side models that co-optimize 
transmission investments and DSO contracts, which will be fully presented in our future work.
\end{remark}

\section{Solution Method and Workflow}
\label{sec:workflow}

In the previous section, constraints~\eqref{eq:M3_deliver} and~\eqref{eq:M4_deliver} require verifying deliverability over all calls in $\Omega_\nu$, all scenarios $s\in\mathcal{S}$, and all windows $\nu\in\mathcal{W}$. The full formulation of the operational and planning feasibility sets $\mathcal{X}^{\rm op}_s$ and $\mathcal{X}^{\rm pl}$, including linearized DistFlow constraints, distributed generation models, and investment cost functions, follows~\cite{11271528}. In this paper, we adopt a \emph{stress-screening} approximation that replaces the deliverability check over all admissible calls in $\Omega_\nu$ with a finite set $\widetilde{\Omega}_\nu\subset\Omega_\nu$ of representative patterns, enabling direct MILP solution; the exact verification procedure is deferred to future work.

\subsection{Stress-Call Screening Set}

For each window $\nu$ with $|\mathcal{T}(\nu)|=H$ steps, 
let $\tau^{\downarrow}_\nu := \lfloor\theta^{\downarrow}_\nu/\Delta t\rfloor$.
Four canonical downward patterns $m\in\mathcal{M}=\{\texttt{base}, 
\texttt{sust}, \texttt{start}, \texttt{end}\}$ are defined as
\begin{equation}
\xi^{\downarrow,m}_t :=
\begin{cases}
    0 & m = \texttt{base}, \\
    \dfrac{\theta^{\downarrow}_\nu R^{\downarrow}_\nu}{H\,\Delta t}
      & m = \texttt{sust}, \\
    R^{\downarrow}_\nu \cdot \mathbf{1}[t \leq t_{\tau^{\downarrow}_\nu}]
      & m = \texttt{start}, \\
    R^{\downarrow}_\nu \cdot \mathbf{1}[t \geq t_{H-\tau^{\downarrow}_\nu+1}]
      & m = \texttt{end},
\end{cases}
\end{equation}
where $\mathbf{1}[\cdot]$ is the indicator function. Upward patterns are defined symmetrically by replacing $\downarrow$ with $\uparrow$. 
Together, these patterns stress the service envelope from complementary power--energy angles: \texttt{base} provides a no-call reference; \texttt{sust} spreads the full energy budget uniformly at reduced power, stressing the energy-to-power ratio of the distribution network's flexibility resources; \texttt{start} and \texttt{end} concentrate full-power calls at the beginning and end of the window, respectively, testing whether the network can sustain delivery after early depletion or enter service with sufficient resource headroom. 
The screening set is thus:
\begin{equation}
    \widetilde{\Omega}_\nu(\boldsymbol{R},\boldsymbol{E}) 
    := \bigl\{(\xi^{\downarrow,m}, \xi^{\uparrow,m'})
       \mid m,m'\in\mathcal{M}\bigr\}
       \cap\, \Omega_\nu(\boldsymbol{R},\boldsymbol{E})
\end{equation}
and deliverability is enforced for each element of 
$\widetilde{\Omega}_\nu$ across all scenarios $s\in\mathcal{S}$ 
and windows $\nu\in\mathcal{W}$, yielding 
$|\widetilde{\Omega}_\nu|\cdot|\mathcal{S}|$ feasibility 
blocks per window per budget tier.



\begin{algorithm}[t]
\small
\DontPrintSemicolon
\SetAlgoLined
\SetKwInOut{Input}{Input}
\SetKwInOut{Output}{Output}
\caption{Temporal NRCC product construction}
\label{alg:temporal_nrcc}
\Input{Feasibility sets $\mathcal{X}^{\rm pl}$,
  $\{\mathcal{X}^{\rm op}_s\}_{s\in\mathcal{S}}$; budget
  tiers $\{\Delta\Gamma_k\}_{k\in\mathcal{K}}$; window
  parameters $(\theta_\nu^{\downarrow/\uparrow}, \rho_\nu,
  \beta_\nu^{\downarrow/\uparrow})_{\nu\in\mathcal{W}}$;
  rebound windows $\mathcal{P}(\nu)$, $\mathcal{R}(\nu)$.}
\Output{Per tier $k$: P0 bounds $(\lambda^D_k,\lambda^R_k)$;
  P1 envelope $(R^{\star\downarrow}_{\nu,k},
  R^{\star\uparrow}_{\nu,k}, E^{\star\downarrow}_{\nu,k},
  E^{\star\uparrow}_{\nu,k})_{\nu\in\mathcal{W}}$;
  P2 rebound bounds $\eta_{k,q}$, $q\in\{{\rm a,b,c}\}$.}
\BlankLine
Solve~\eqref{eq:M1}; record $\Gamma^0$ and
  $\bar{p}^0_{t,s}$, $\forall t,s$\;
\For{$k\in\mathcal{K}$}{
  \nonl\textit{P0: atemporal NRCC}\;
  Solve~\eqref{eq:M2} at $\Gamma^0\!+\!\Delta\Gamma_k$;
    record $\lambda^D_k$, $\lambda^R_k$\;
  \nonl\textit{P1: service-window envelope}\;
  Build $\widetilde{\Omega}_\nu$ via four stress patterns,
    $\forall\nu\in\mathcal{W}$\;
  Solve~\eqref{eq:M3} over $\widetilde{\Omega}_\nu\times
    \mathcal{S}$; record
    $(R^{\star\downarrow}_{\nu,k}, R^{\star\uparrow}_{\nu,k},
    E^{\star\downarrow}_{\nu,k}, E^{\star\uparrow}_{\nu,k})$,
    $\forall\nu$\;
  \nonl\textit{P2: rebound-aware refinement}\;
  \For{$q\in\{{\rm a,b,c}\}$}{
    Solve~\eqref{eq:M4}-$q$ with
      floors~\eqref{eq:M4_lock}; record $\eta_{k,q}$\;
  }
}
\end{algorithm}

\subsection{Product Construction Procedure}

Algorithm~\ref{alg:temporal_nrcc} constructs the full product hierarchy by sweeping the budget levels $\{\Delta\Gamma_k\}$ and solving, at each tier, Models~1--4 
in sequence.

\subsubsection{Baseline and P0 (Models~1--2)}
Model~1 is solved once to obtain the least-cost budget $\Gamma^0$ and the baseline netload profiles $\bar{p}^0_{t,s}$. For each tier $k$, Model~2 is then solved at $\Gamma^0+\Delta\Gamma_k$ to obtain the P0 bounds $(\lambda^D_k, \lambda^R_k)$, which serve as the outside-window guarantees in all subsequent stages.

\subsubsection{P1 Construction (Model~3)}
For each tier $\Delta\Gamma_k$, the deliverability constraint~\eqref{eq:M3_deliver} is replaced by finite feasibility blocks over $\widetilde{\Omega}_\nu\times\mathcal{S}$, appended directly to the investment optimization as a single MILP. The solution yields the per-window service envelope $(R^{\star\downarrow}_{\nu,k}, E^{\star\downarrow}_{\nu,k}, R^{\star\uparrow}_{\nu,k}, E^{\star\uparrow}_{\nu,k})_{\nu\in\mathcal{W}}$, which serves as the P1-capacity floor~\eqref{eq:M4_lock} for P2.

\subsubsection{P2 Construction (Model~4)}
Given the P1-optimal envelope, Model~4 is solved once per variant $q\in\{{\rm a,b,c}\}$. The P1-capacity floor~\eqref{eq:M4_lock} pins each window's deliverable adjustments and energy budgets at their P1-optimal values, reducing the second-stage problem to minimizing $\eta$ under the chosen outside-window governance rule~\eqref{eq:M4_outa} or~\eqref{eq:M4_outb}.
\section{Case Study and Results}
\label{sec:results}

\subsection{Test System and Setup}

The framework is applied to a representative feeder (\texttt{p8udt1252-p8uhs6\_1247x}, located in San Francisco, CA) from the SMART-DS synthetic distribution dataset \cite{nrel2020smartds}. The feeder has 65 buses and 60 lines, and has 5/5/1 candidate lines/storage units/voltage regulators for reinforcement or installation, respectively. Parameters and costs associated with candidate
line reinforcements, storage systems, and VRs on the distribution network side are extracted from \cite{kersting2018distribution, nrel2019costdatabase}. 
Long-term netload uncertainty is generated based on the same procedure in \cite{11271528} and reduced to 3 scenarios covering 3 representative days. Combined with the four canonical stress-call patterns from Algorithm~\ref{alg:temporal_nrcc}, this yields 12 feasibility blocks per window per budget level. For brevity, the case study considers a single 
downward service window ($R^\uparrow = 0$); 
upward service windows are left for future work.

The service window $\mathcal{T}(\nu)$ covers hours~16--19, when the evening demand ramp after distributed generation subsides produces the highest and most temporally concentrated netload across all scenarios, making it the most likely target for TSO-initiated flexibility calls and the most demanding test of P1 deliverability. The protected window $\mathcal{P}(\nu)$ (hours~20--23) immediately follows service, and the designated rebound window $\mathcal{R}(\nu)$ (hours~0--5) targets the overnight low-stress slot, together covering the full post-service recovery cycle that P2 governs. Table~\ref{tab:setup} summarizes the remaining parameters.

\begin{table}[t]
\centering
\caption{Case study parameters.}
\label{tab:setup}
\small
\begin{tabular}{@{}ll@{}}
\toprule
\textbf{Parameter} & \textbf{Value} \\
\midrule
Candidate storage duration         & 2 h \\
Service window $\mathcal{T}(\nu)$   & Hours 16--18 \\
Duration multiplier $\theta^\downarrow$ & 2 h \\
P2-a protected window $\mathcal{P}(\nu)$ & Hours 19--22 \\
P2-b rebound window $\mathcal{R}(\nu)$   & Hours 0--5  \\
P2-c rebound window & $\mathcal{T}\setminus\mathcal{T}(\nu)$  \\
\bottomrule
\end{tabular}
\end{table}

\begin{figure}[t]
\centering
\begin{subfigure}[b]{0.49\columnwidth}
  \centering
  \includegraphics[width=\linewidth]{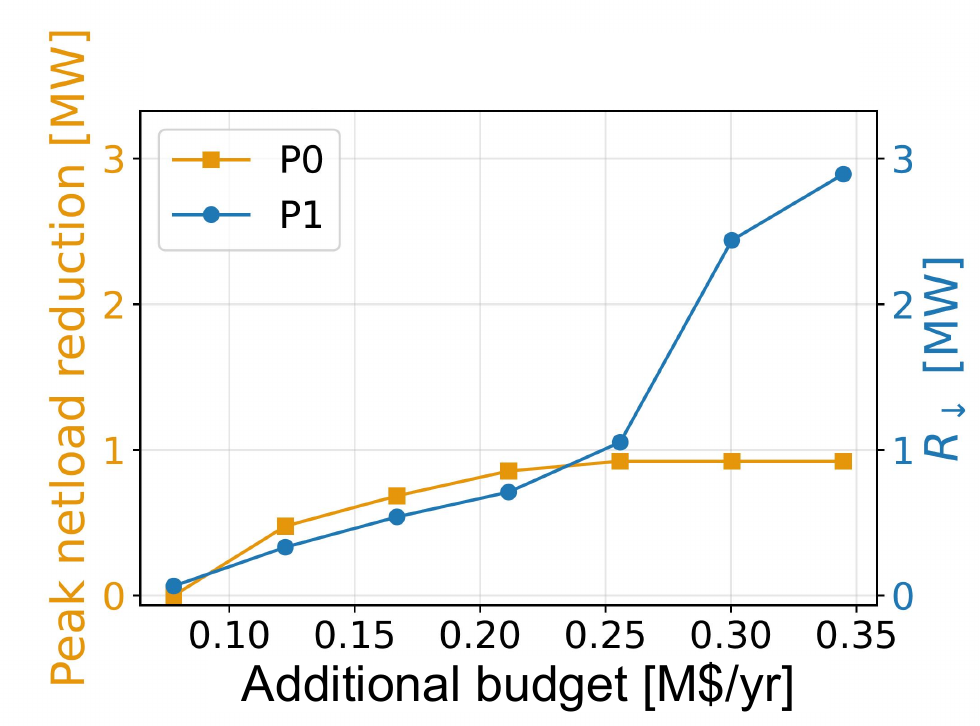}
  \caption{\scriptsize P0 peak cap and P1 $R^{\downarrow}$.}
  \label{fig:p0_p1}
\end{subfigure}\hfill
\begin{subfigure}[b]{0.48\columnwidth}
  \centering
  \includegraphics[width=\linewidth]{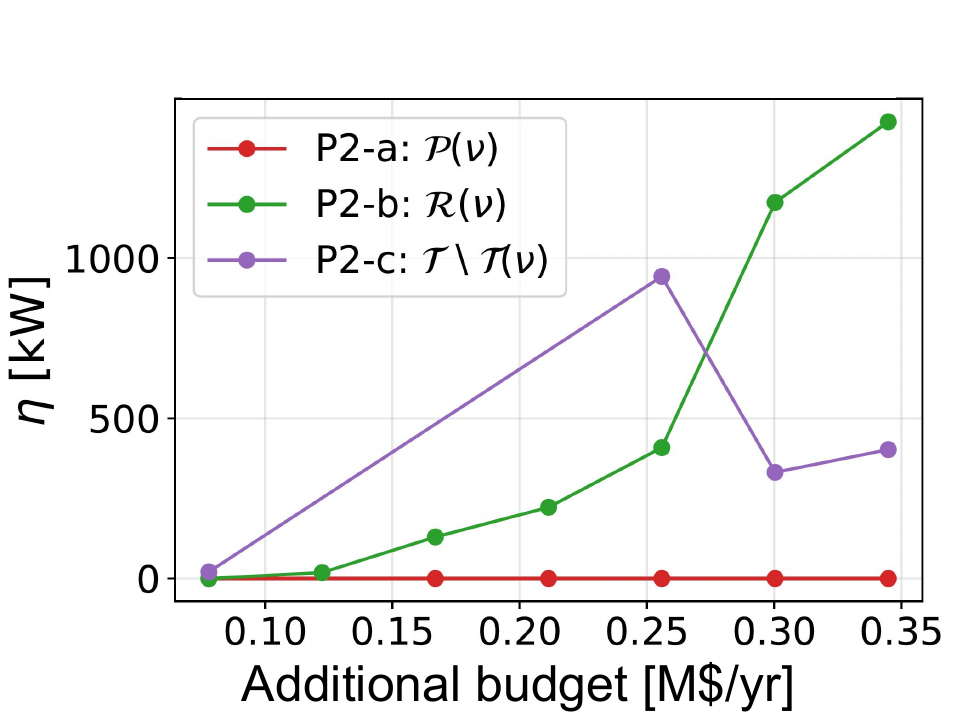}
  \caption{\scriptsize P2 rebound $\eta$ vs.\ budget.}
  \label{fig:p2_eta}
\end{subfigure}
\caption{P0 peak netload reduction, P1 service capability, and P2 rebound metric across budget tiers.}
\label{fig:budget_sweep}
\end{figure}

\begin{figure}[t]
\centering
\includegraphics[width=\columnwidth]{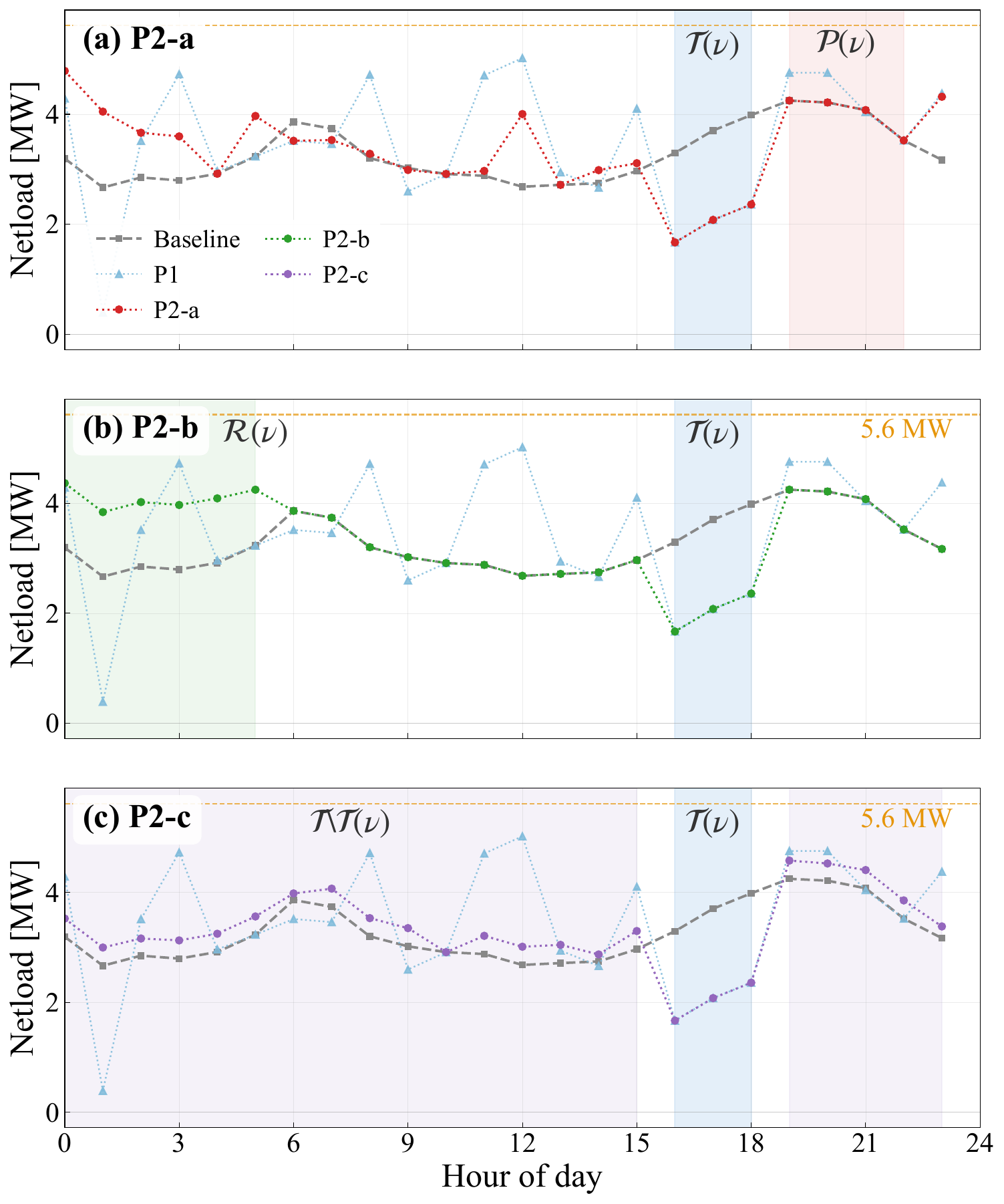}
\caption{Substation netload profiles under sustained down-call for P2 variants at reference tier. Shaded regions mark service window $\mathcal{T}(\nu)$ and variant-specific recovery windows.}
\label{fig:p2_variants}
\end{figure}

\subsection{P0 and P1 Results}

Fig.~\ref{fig:budget_sweep}(a) reports the P0 peak cap $\lambda^D_k$ and P1 downward rating $R^\downarrow$ versus incremental budget $\Delta\Gamma$. The cost-optimal baseline yields $\Gamma^0\approx\$77{,}859$/yr. P0 peak reduction saturates beyond $\Delta\Gamma\approx 178$k/yr, with the cap declining from 6{,}537~kW to 5{,}615~kW ($\sim$14\%) as scenario-driven netload uncertainty is progressively eliminated. In contrast, $R^\downarrow$ grows throughout, reaching $\sim$2{,}900~kW at the highest tier, driven by storage investment. The divergence of the two curves beyond saturation reveals that budget increments yielding negligible peak reduction can still unlock substantial temporal flexibility, which is a value that the atemporal P0 interface alone cannot capture. 

Two structural features of the P1 curve merit attention. First, $R^\downarrow$ grows approximately linearly up to $\Delta\Gamma\approx 134$k/yr ($R^\downarrow\approx 710$~kW), consistent with storage-dominated investment at a roughly constant marginal cost. Second, a pronounced jump occurs near $\Delta\Gamma\approx 222$k/yr, where $R^\downarrow$ rises steeply from ${\sim}1{,}050$~kW to ${\sim}2{,}440$~kW. This step change reflects a discrete investment regime transition, where additional storage capacity becomes accessible only after upstream line reinforcements are activated, and underscores that flexibility supply is not a smooth function of budget on networks with discrete investment options.

\subsection{P2 Results: Rebound Response}

Fig.~\ref{fig:budget_sweep}(b) shows $\eta$ versus 
$\Delta\Gamma$ for all three P2 variants. In addition 
to P2-a and P2-b, we evaluate P2-c, a global rebound 
variant that bounds deviation from baseline over all 
non-service hours $\mathcal{T}\setminus\mathcal{T}(\nu)$ 
without designating a specific recovery window. The 
three variants exhibit qualitatively different budget 
responses, showing that governance design rather than 
budget allocation is the primary determinant of rebound 
scaling behavior.

\textbf{P2-a} maintains $\eta=0$ across all budget levels. Under the P2-a governance rule, storage is not required to complete recharge within $\mathcal{P}(\nu)$; hours outside $\mathcal{T}(\nu)\cup\mathcal{P}(\nu)$ remain subject only to the P0 peak cap, providing ample time for storage to recharge without deviating from baseline during the protected window. Consequently, $\mathcal{P}(\nu)$ experiences zero netload deviation regardless of how much energy was discharged during service, and $\eta=0$ holds even at the highest tier where $R^\downarrow\approx 2{,}900$~kW.

\textbf{P2-b} shows monotonically increasing $\eta$. As storage investment enables greater $R^\downarrow$, more energy must be recovered within the overnight slot $\mathcal{R}(\nu)$ (hours~0--5), progressively saturating its limited headroom. At the highest budget tier, $\eta$ exceeds 1{,}400~kW, revealing a fundamental tension in the P2-b design: stronger service capability directly worsens rebound concentration.

\textbf{P2-c} exhibits non-monotone behavior, with $\eta$ peaking near ${\sim}940$~kW at intermediate budgets before declining. The initial rise parallels P2-b as growing service energy increases total recovery volume. At higher budgets, however, expanded storage and network capacity enable more effective temporal spreading, reducing the worst-case per-hour deviation.



\subsection{Load Profile Analysis}

Fig.~\ref{fig:p2_variants} overlays representative substation netload profiles for the sustained down-call scenario. P1 is included as a reference: while it delivers the same service-window down-ramp as all P2 variants, its outside-window profile fluctuates freely as storage recharges without any rebound governance, producing large uncontrolled deviations from baseline. The P2 variants each suppress this behavior differently.

Under \textbf{P2-a}, the profile within $\mathcal{P}(\nu)$ tracks baseline closely, consistent with $\eta=0$, while hours outside $\mathcal{T}(\nu)\cup\mathcal{P}(\nu)$ retain P1-level freedom below the peak cap. The TSO observes intended service followed by seamless return to baseline in the protected window. Under \textbf{P2-b}, all hours outside $\mathcal{T}(\nu)\cup\mathcal{R}(\nu)$ adhere strictly to baseline, eliminating the uncontrolled fluctuations visible in P1 everywhere except the overnight slot (hours~0--5), where energy recovery is visibly concentrated. This sharp concentration directly explains the monotonically growing $\eta$: the entire post-service energy debt is compressed into a window whose headroom does not scale with investment. Under \textbf{P2-c}, the profile tracks baseline throughout all non-service hours with small uniform deviations, substantially smoothing the P1 fluctuations without designating a specific recovery slot.

The three variants thus span a flexibility--predictability spectrum. The choice among them depends on whether the TSO prioritizes zero post-service deviation (P2-a), full-day baseline adherence outside a designated slot (P2-b), or global worst-case deviation minimization (P2-c).

\subsection{Discussion}

The results reveal three implications for temporal flexibility product design. First, the P0 and P1 curves diverge at higher budget tiers, nonetheless, confirming that incremental investments in storage can be better valued by time-window-based products by considering temporal flexibility. Second, rebound behavior is governed by contract design rather than budget: despite sharing the same P1-optimal service capacity, the three P2 variants exhibit qualitatively distinct $\eta$-budget relationships, confirming that a rebound bound is meaningful only when paired with its governance structure. Third, the choice among P2 variants involves a fundamental trade-off: P2-a offers the strongest rebound guarantee with moderate outside-window freedom, P2-b provides full baseline predictability at the cost of growing rebound concentration, and P2-c minimizes peak hourly deviation globally without protected or rebound window-specific commitments. The appropriate variant depends on the TSO's operational priorities.

\section{Conclusion}
\label{sec:conclusion}

This paper proposed a hierarchy of certified temporal flexibility products (P0, P1, P2) within the NRCC framework for TSO--DSO coordinated planning. P1 maps distribution investment into scenario-robust service-window envelopes; P2 adds lexicographic rebound minimization with three governance variants that exhibit qualitatively distinct budget--rebound responses, confirming that rebound guarantees are meaningful only when specified jointly with governance rules.

Future work will replace the stress-screening approximation with the full deliverability verification procedure for stronger robustness guarantees, integrate the P1 and P2 products into the TSO-side portfolio selection model to quantify the transmission planning value of rebound-certified flexibility, and validate the proposed framework across multi-substation feeders with diverse load shapes to characterize the conditions under which each governance variant is most advantageous.

\bibliographystyle{IEEEtran}
\bibliography{reference}

\end{document}